\documentclass[fleqn,12pt]{iopart}

\usepackage{iopams}  
\usepackage{graphicx}
\usepackage{multirow}
\usepackage{amssymb}
\expandafter\let\csname equation*\endcsname\relax
\expandafter\let\csname endequation*\endcsname\relax
\usepackage{amsmath}
\usepackage{microtype}
\usepackage{url}
\usepackage{natbib}
\usepackage{appendix}
\usepackage[margin=1cm]{caption}
\usepackage{braket}
\usepackage{fullpage}
\usepackage{hyperref}
\usepackage{color}

\newcommand{\Msun}{\ensuremath{M_{ \odot }}}
\newcommand{\degree}{\ensuremath{^\circ}}
\newcommand{\innerprod}[2]{\ensuremath{\left({#1}\middle|{#2}\right)}}

\begin{document}

\title{Gravitational-wave sensitivity curves}
\author{C J Moore$^1$, R H Cole$^1$ \& C P L Berry$^{1,\,2}$}
\address{$^1$ Institute of Astronomy, Madingley Road, Cambridge, CB3 0HA, United Kingdom}
\address{$^2$ School of Physics and Astronomy, University of Birmingham, Edgbaston, Birmingham B15 2TT, United Kingdom}
\ead{cjm96@cam.ac.uk}

\begin{abstract}
There are several common conventions in use by the gravitational-wave community to describe the amplitude of sources and the sensitivity of detectors. These are frequently confused. We outline the merits of and differences between the various quantities used for parameterizing noise curves and characterizing gravitational-wave amplitudes. We conclude by producing plots that consistently compare different detectors. Similar figures can be generated on-line for general use at \url{http://rhcole.com/apps/GWplotter}.
\end{abstract}

\pacs{04.30.--w, 04.30.Db, 04.80.Nn, 95.55.Ym}

\section{Introduction}

The next few years  promise to deliver the first direct detection of gravitational waves (GWs). This will most likely be achieved by the advanced versions of the Laser Interferometer Gravitational-wave Observatory \citep[LIGO;][]{2010CQGra..27h4006H} and Virgo \citep{Acernese2009} detectors operating in the frequency range $(10$--$10^{3})~\mathrm{Hz}$. By the end of the decade, it is expected that pulsar timing arrays \citep[PTAs;][]{1990ApJ361300F} will also detect very low frequency GWs around $10^{-8}~\mathrm{Hz}$. Further into the future, space-based detectors, such as the evolving Laser Interferometer Space Antenna \citep[eLISA;][]{TheGravitationalUniverse}
\footnote{The e in eLISA originally stood for evolved, but the preference now is for evolving.}
These advances shall herald the beginning of multi-wavelength GW astronomy as a means of observing the Universe.

There already exists an extensive literature assessing the potential of all of these detectors to probe the astrophysics of various sources. There are several different methods commonly used to describe the sensitivity of a GW detector and the strength of a GW source. It is common practice to summarise this information on a sensitivity-curve plot. When producing these plots, it is desirable to have a consistent convention between detectors and sources that allows information about both to be plotted on the same graph. Ideally, the detectors and sources are represented in such a way that the relative detectability of the signals is immediately apparent.

In this work, we discuss the differing conventions commonly used in GW astronomy. The amplitude of a GW is a strain, a dimensionless quantity $h$. This gives a fractional change in length, or equivalently light travel time, across a detector. The strain is small, making it a challenge to measure: we are yet to obtain a direct detection of a GW. To calibrate our expectations for future detections it is necessary to quantify the sensitivity of our instruments and the strength of their target signals. When discussing the loudness of sources and the sensitivity of detectors there are three commonly used parametrizations based upon the strain: the characteristic strain, the power spectral density (PSD) and the spectral energy density. We aim to disambiguate these three and give a concrete comparison of different detectors. It is hoped that this will provide a useful reference for new and experienced researchers in this field alike. 

We begin by expounding the various conventions and the relationships between the conventions in sections \ref{sec:conventions} and \ref{sec:voc}. A review of GW detectors (both current and proposed) is given in section \ref{sec:detectors} and a review of GW sources is given in section \ref{sec:sources}. In \ref{app:a}, several example sensitivity curves are presented. A website where similar figures can be generated is available at \url{http://rhcole.com/apps/GWplotter}. Here, the user may select which sources and detectors to include to tailor the figure to their specific requirements.

\section{Signal parametrization}\label{sec:conventions}

\subsection{Signal analysis preliminaries}

Gravitational radiation has two independent polarization states denoted $+$ and $\times$; a general signal can be described as a linear combination of the two polarization states, $h \nobreak=\nobreak A_{+} h_{+}\nobreak +\nobreak A_{\times} h_{\times}$. The sensitivity of a detector to these depends upon the relative orientations of the source and detector. The output of a gravitational wave detector $s(t)$ contains a superposition of noise $n(t)$ and (possibly) a signal $h(t)$,
\begin{equation}
s(t) = n(t)+h(t) \; .
\end{equation}

We shall have recourse to work with the Fourier transform of the signal, using the conventions that
\begin{eqnarray} \label{eq:defFourier}
\tilde{x}(f) &= {\mathcal{F}}\left\{x(t)\right\}(f) &= \int_{-\infty}^{\infty}\mathrm{d}t\;x(t)\exp (-2\pi \rmi f t ) \; , \\
x(t) &= {\mathcal{F}}^{-1}\left\{\tilde{x}(f)\right\}(t) &= \int_{-\infty}^{\infty}\mathrm{d}f\;\tilde{x}(f)\exp (2\pi \rmi f t ) \; .
\end{eqnarray}

For simplicity, it is assumed that the noise in the GW detector is stationary and Gaussian (with zero mean); under these assumptions the noise is fully characterised via the one-sided noise PSD $S_{n}(f)$, 
\begin{equation}\label{eq:psd}
\left<\tilde{n}(f)\tilde{n}^{*}(f')\right>=\frac{1}{2}\delta (f-f')S_{n}(f) \; ,
\end{equation}
where angle brackets $\left<\ldots\right>$ denote an ensemble average over many noise realisations \citep{Cutler1994}. In reality, we have only a single realisation to work with, but the ensemble average can be replaced by a time average for stationary stochastic noise. The procedure is to measure the noise over a sufficiently long duration $T$ and then compute the Fourier transform $\tilde{n}(f)$ with a frequency resolution $\Delta f = T^{-1}$; this is repeated many times to give an average. The noise PSD $S_{n}(f)$ has units of inverse frequency.

Since the GW signal and detector output are both real, it follows that $\tilde{h}(-f)=\tilde{h}^{*}(f)$ and $\tilde{n}(-f)=\tilde{n}^{*}(f)$; therefore, $S_{n}(f)=S_{n}(-f)$. The fact that $S_{n}(f)$ is an even function means that Fourier integrals over all frequencies can instead be written as integrals over positive frequencies only, e.g., (\ref{eq:meansquare2}) and (\ref{eq:snrinnerprod}); it is for this reason that $S_{n}(f)$ is called the \emph{one-sided} PSD.\footnote{An alternative convention is to use the two-sided PSD $S^{(2)}_{n}(f) = S_{n}(f)/2$.}

When integrated over all positive frequencies, the PSD gives the mean square noise amplitude. Starting by taking the time average of the square of the detector noise:
\begin{eqnarray}
\overline{\left|n(t)\right|^{2}} &= \lim_{T\rightarrow\infty}\frac{1}{2T} & \int_{-T}^{T}\mathrm{d}t\; n(t)n^{*}(t) \\
 &= \lim_{T\rightarrow\infty}\frac{1}{2T} & \int_{-T}^{T}\mathrm{d}t\; \int_{-\infty}^{\infty}\mathrm{d}f\;\int_{-\infty}^{\infty}\mathrm{d}f'\;\tilde{n}(f)\tilde{n}^{*}(f')\exp\left(2\pi \rmi ft\right)\exp\left(-2\pi \rmi f't\right) \nonumber\\
 &= \lim_{T\rightarrow\infty}\frac{1}{2T} & \int_{-T}^{T}\mathrm{d}t\; \int_{-\infty}^{\infty}\mathrm{d}f\;\int_{-\infty}^{\infty}\mathrm{d}f'\;{\mathcal{F}}\left\{ n(\tau) \right\}(f)\left[{\mathcal{F}}\left\{ n(\tau) \right\}(f')\right]^{*} \nonumber \\*
\label{eq:meansquare} & & \times \vphantom{\int_{-T}^{T}} \exp\left(2\pi \rmi ft\right)\exp\left(-2\pi \rmi f't\right)\;,
\end{eqnarray}
where we have substituted in using the definitions of the Fourier transform and its inverse. A property of Fourier transforms is that a time-domain translation by amount $t$ is equivalent to a frequency-domain phase change $2\pi ft$; if ${\mathcal{F}}\left\{ n(\tau) \right\}(f) = \tilde{n}(f)$, then ${\mathcal{F}}\left\{ n(\tau-t) \right\}(f) = \tilde{n}(f)\exp (2\pi \rmi ft )$. Therefore, the exponential factors in (\ref{eq:meansquare}) may be absorbed as
\begin{eqnarray}
\overline{\left|n(t)\right|^{2}} &=& \lim_{T\rightarrow\infty}\frac{1}{2T}\int_{-T}^{T} \mathrm{d}t\:\int_{-\infty}^{\infty}\mathrm{d}f\;\int_{-\infty}^{\infty}\mathrm{d}f'\;{\mathcal{F}}\left\{ n(\tau-t) \right\}(f)\left[{\mathcal{F}}\left\{ n(\tau-t) \right\}(f')\right]^{*} \, .
\end{eqnarray}
Since the noise is a randomly varying signal, we can use the ergodic principle to equate a time average, denoted by $\overline{\left(\ldots\right)}$, with an ensemble average, denoted by $\left<\ldots\right>$. The noise is stationary, consequently, its expectation value is unchanged by the time-translation performed above. Therefore, using (\ref{eq:psd}), the mean square noise amplitude is given by
\begin{eqnarray}\label{eq:meansquare1}
\overline{\left|n(t)\right|^{2}} &=& \int_{-\infty}^{\infty}\mathrm{d}f\;\int_{-\infty}^{\infty}\mathrm{d}f'\;\left<\tilde{n}(f)\tilde{n}^{*}(f')\right> \\
&=& \int_{-\infty}^{\infty}\mathrm{d}f\;\int_{-\infty}^{\infty}\mathrm{d}f'\;\frac{1}{2}S_{n}(f)\delta(f-f') \nonumber \\
&=& \int_{0}^{\infty}\mathrm{d}f\; S_{n}(f) \;.
\label{eq:meansquare2}
\end{eqnarray}

Given a detector output, the challenge is to extract the signal. There is a well known solution to this problem that involves constructing a Wiener optimal filter \citep{Wiener49}. Let $K(t)$ be a real filter function with Fourier transform $\tilde{K}(f)$. Convolving this with the detector output gives a contribution from the signal and a contribution from the noise,
\begin{equation}\label{eq:conv}
\left(s*K\right)(\tau) = \int_{-\infty}^{\infty}\mathrm{d}t\;\left[h(t)+n(t)\right]K(t-\tau) \approx \mathcal{S} + \mathcal{N} \; .
\end{equation}
The signal contribution $\mathcal{S}$ is defined as the expectation of the convolution in (\ref{eq:conv}) when a signal is present, maximised by varying the offset to achieve the best overlap with the data. Since the expectation of pure noise is zero it follows that
\begin{equation}
\mathcal{S} = \int_{-\infty}^{\infty}\mathrm{d}t\;h(t)K(t)=\int_{-\infty}^{\infty}\mathrm{d}t\;h(t)K^{*}(t)=\int_{-\infty}^{\infty}\mathrm{d}f\; \tilde{h}(f)\tilde{K}^{*}(f) \; .
\end{equation}
The squared contribution from noise $\mathcal{N}^{2}$ is defined as the mean square of the convolution in (\ref{eq:conv}) when no signal is present,
\begin{eqnarray} 
\mathcal{N}^{2} &= \int_{-\infty}^{\infty}\mathrm{d}t\;\int_{-\infty}^{\infty}\mathrm{d}t'\;K(t)K(t')\left<n(t)n(t')\right> \nonumber \\
 &= \int_{-\infty}^{\infty}\mathrm{d}t\;\int_{-\infty}^{\infty}\mathrm{d}t'\;K(t)K^{*}(t')\int_{-\infty}^{\infty}\mathrm{d}f\;\int_{-\infty}^{\infty}\mathrm{d}f'\;\left<\tilde{n}(f)\tilde{n}^{*}(f')\right>\exp\left[2\pi\rmi(ft-f't')\right]\nonumber \\
 &= \int_{-\infty}^{\infty}\mathrm{d}f \; \frac{1}{2}S_{n}(f)\tilde{K}(f)\tilde{K}^{*}(f) \; ,
 \end{eqnarray}
using the definition of $S_{n}(f)$ from (\ref{eq:psd}). Hence the signal-to-noise ratio (SNR) $\varrho$ is given by
\begin{equation}\label{eq:SNRinnerprod} 
\varrho^{2} = \frac{\mathcal{S}^{2}}{\mathcal{N}^{2}}= \frac{\innerprod{\frac{1}{2}S_{n}(f)\tilde{K}(f)}{\tilde{h}(f)}^{2}}{\innerprod{\frac{1}{2}S_{n}(f)\tilde{K}(f)}{\frac{1}{2}S_{n}(f)\tilde{K}(f)}},
\end{equation}
where we have introduced the inner product between signal $\tilde{A}$ and $\tilde{B}$ as \citep{Finn1992}
\begin{equation}\label{eq:snrinnerprod} \innerprod{\tilde{A}(f)}{\tilde{B}(f)} = 4\Re\left\{\int_{0}^{\infty}\mathrm{d}f\;\frac{\tilde{A}^{*}(f)\tilde{B}(f)}{S_{n}(f)}\right\} \; .\end{equation}
The optimum filter is that function $\tilde{K}(f)$ which maximises the SNR in (\ref{eq:SNRinnerprod}). From the Cauchy--Schwarz inequality, it follows that the optimum filter is
\begin{equation}
\tilde{K}(f)=\frac{\tilde{h}(f)}{S_{n}(f)} \; .
\end{equation}
This is the Wiener filter, which may be multiplied by an arbitrary constant since this does not change the SNR. Using this form for $\tilde{K}(f)$, the squared SNR is
\begin{equation}
\varrho^2 = \int_0^\infty\mathrm{d}f \frac{4| \tilde{h}(f)|^{2}}{S_n(f)} = \innerprod{\tilde{h}(f)}{\tilde{h}(f)}.
\label{eq:traditionalSNR} 
\end{equation}
In order to construct the Wiener filter, it is necessary to know \emph{a priori} the form of the signal, $\tilde{h}(f)$, for this reason the Wiener filter is sometimes called the \emph{matched} filter.

Whilst the magnitude of the Fourier transform of the signal $|\tilde{h}(f)|$ provides a simple quantification of the GW amplitude as a function of frequency, it has one main deficiency. For an inspiralling source, the instantaneous amplitude can be orders of magnitude below the noise level in a detector; however, as the signal continues over many orbits, the SNR can be integrated up to a detectable level. It is useful to have a quantification of the GW amplitude that accounts for this effect; we shall now describe how this can be achieved.

\subsection{Characteristic strain}\label{sec:character-strain}

The characteristic strain $h_\mathrm{c}$ is designed to include the effect of integrating an inspiralling signal. Its counterpart for describing noise is the noise amplitude $h_n$. These are defined as
\begin{eqnarray}\label{eq:strain-hc} 
\left[h_\mathrm{c}(f)\right]^{2} &= 4f^{2}\left| \tilde{h}(f) \right|^{2} \; ,\\
\left[h_{n}(f)\right]^{2} &= fS_{n}(f) \; ,
\label{eq:strain-hn}
\end{eqnarray}
such that the SNR in (\ref{eq:traditionalSNR}) may be written
\begin{equation}\label{eq:hc} 
\varrho^{2} = \int_{-\infty}^{\infty} \mathrm{d}\left(\log f\right)\; \left[\frac{h_\mathrm{c}(f)}{h_{n}(f)}\right]^{2} \;.
\end{equation}
The strain amplitudes $h_\mathrm{c}(f)$ and $h_{n}(f)$ are dimensionless. Using this convention, when plotting on a log--log scale, the area between the source and detector curves is related to the SNR via (\ref{eq:hc}). This convention allows the reader to integrate by eye to assess the detectability of a given source (see figure \ref{fig:hc}).

An additional advantage of this convention is that the values on the strain axis for the detector curve $h_n(f)$ have a simple physical interpretation: they correspond to the root-mean-square noise in a bandwidth $f$. One downside to plotting characteristic strain is that the values on the strain axis $h_\mathrm{c}(f)$ do not directly relate to the amplitude of the waves from the source. 
Another disadvantage is that applying (\ref{eq:strain-hc}) to a monochromatic source gives a formally undefined answer. The correct identification of characteristic strain for a monochromatic source is the amplitude of the wave times the square root of the number of periods observed (see section \ref{sec:insp}).

\subsection{Power spectral density}\label{sec:psd}

A second commonly used quantity for sensitivity curves is the square root of the PSD or the amplitude spectral density (see figure \ref{fig:S}). When discussing a detector, rearranging (\ref{eq:strain-hn}) gives
\begin{equation}\label{eq:temp1}
\sqrt{S_{n}(f)} = h_{n}(f)f^{-1/2} \; ;
\end{equation}
by analogy, we can define an equivalent for source amplitudes
\begin{equation}
\sqrt{S_{h}(f)} = h_\mathrm{c}(f)f^{-1/2} = 2 f^{1/2} \left| \tilde{h}(f) \right| \; ,
\label{eq:ShforSources}
\end{equation}
where we have used (\ref{eq:strain-hc}). Both $\sqrt{S_{n}(f)}$ and $\sqrt{S_{h}(f)}$ have units of $\mathrm{Hz^{-1/2}}$. The root PSD is the most frequently plotted quantity in the literature.

The PSD, as defined by (\ref{eq:psd}), has the nice property, demonstrated in (\ref{eq:meansquare2}), that integrated over all positive frequencies it gives the mean square amplitude of the noise in the detector. However, in one important regard it is less appealing than characteristic strain: the height of the source above the detector curve is no longer directly related to the SNR.

\subsection{Energy density}\label{sec:energy-density}

A third way of describing the amplitude of a GW is through the energy carried by the radiation. This has the advantage of having a clear physical significance. The energy density is most commonly used in sensitivity curves showing stochastic backgrounds of GWs (see section \ref{sec:stoch}).

The energy in GWs is described by the Isaacson stress--energy tensor \citep[section 35.15]{MTW}
\begin{equation}
T_{\mu\nu}=\frac{c^{4}}{32\pi G}\left<\partial_{\mu}\bar{h}_{\alpha\beta}\partial_{\nu}\bar{h}^{\alpha\beta}\right> \;,
\end{equation}
where the angle brackets denote averaging over several wavelengths or periods, and $\bar{h}_{\alpha\beta}$ is the transverse-traceless metric perturbation. The energy density $\rho c^{2}$ is given by the $T_{00}$ component of this tensor. Consequently \citep[cf.][]{Berry2013},
\begin{eqnarray}
\rho c^{2} &=& \frac{c^{2}}{16\pi G}\int_{-\infty}^{\infty}\mathrm{d}f\;\left(2\pi f\right)^{2}\tilde{h}(f)\tilde{h}^{*}(f) \\*
 &=& \int_{0}^{\infty}\mathrm{d}f\;\frac{\pi c^{2}}{4G}f^{2}S_{h}(f)\; , \label{eq:specNRGdensity}
\end{eqnarray} 
where the definition (\ref{eq:ShforSources}) has been used. The integrand in (\ref{eq:specNRGdensity}) is defined as the spectral energy density, the energy per unit volume of space, per unit frequency \citep{HellingsDowns}
\begin{equation}\label{eq:spectralenergydensity}
S_{\mathrm{E}}(f)=\frac{\pi c^{2}}{4G} f^{2}S_{h}(f) \; ;
\end{equation}
a corresponding expression for the noise can be formulated by replacing $S_h(f)$ with $S_{n}(f)$.

Cosmological studies often work in terms of the dimensionless quantity $\Omega_{\mathrm{GW}}$, the energy density per logarithmic frequency interval, normalised to the critical density of the Universe $\rho_{\mathrm{c}}$,
\begin{equation}
\label{eq:omega}
\Omega_\mathrm{GW}(f) = \frac{fS_{\mathrm{E}}(f)}{\rho_{\mathrm{c}}c^{2}} \; .
\end{equation}
The critical density is
\begin{equation}
\label{eq:crit-density}
\rho_{\mathrm{c}}=\frac{3H_{0}^{2}}{8\pi G} \;,
\end{equation}
where $H_{0}$ is the Hubble constant, commonly parametrized as
\begin{equation}
H_0 = h_{100}\times 100~\mathrm{km\,s^{-1}\,Mpc^{-1}}.
\end{equation}
The reduced Hubble parameter $h_{100}$ has nothing to do with strain. The most common quantity related to energy density to be plotted on sensitivity curves is $\Omega_{\mathrm{GW}}h_{100}^{2}$ (figure \ref{fig:omega}) as this removes sensitivity to the (historically uncertain) measured value of the Hubble constant.

This quantity has one aesthetic advantage over the others: it automatically accounts for there being less energy in low frequency waves of the same amplitude. However, unlike characteristic strain, the area between the source and detector curves is no longer simply related to the SNR.

\subsection{Relating the descriptions}

The dimensionless energy density in GWs $\Omega_{\mathrm{GW}}$, spectral energy density $S_{\mathrm{E}}$, one-sided PSD $S_{h}$, characteristic strain $h_\mathrm{c}$ and frequency-domain strain $\tilde{h}(f)$ are related via
\begin{equation}\label{eq:differentdescriptions}
H_0^2\Omega_\mathrm{GW}(f)= \frac{8 \pi G}{3 c^{2}} fS_{\mathrm{E}}(f) = \frac{2\pi^2}{3} f^3 S_h(f) = \frac{2\pi^2}{3} f^2 \left[h_\mathrm{c}(f)\right]^2 = \frac{8\pi^2}{3} f^4 \left|\tilde{h}(f)\right|^2\; ,
\end{equation}
using (\ref{eq:strain-hc}), (\ref{eq:ShforSources}), (\ref{eq:spectralenergydensity}), (\ref{eq:omega}) and (\ref{eq:crit-density}).
Corresponding expressions for the noise are obtained by substituting $S_{n}(f)$ for $S_h(f)$, $h_{n}(f)$ for $h_\mathrm{c}(f)$ and $\tilde{n}(f)$ for $\tilde{h}(f)$.

\section{Types of source}\label{sec:voc}

GW signals can be broadly split into three categories: those from well-modelled sources, for which we have a description of the expected waveform; stochastic backgrounds, for which we can describe the statistical behaviour; and unmodelled (or poorly-modelled) transient sources. The classic example of a well-modelled source is the inspiral of two compact objects, this is discussed in section \ref{sec:insp}. Stochastic backgrounds can either be formed from many overlapping sources, which could be modelled individually, or from some intrinsically random process, these are discussed in section \ref{sec:stoch}. An example of an unmodelled (or poorly-modelled) transient source is a supernova; searches for signals of this type are often called burst searches and are discussed in section \ref{sec:bursts}.

\subsection{Inspirals}\label{sec:insp}

Inspiralling binaries may the most important GW source. They spend a variable amount of time in each frequency band. If $\phi$ is the orbital phase, then the number of cycles generated at frequency $f$ can be estimated as
\begin{equation}\label{eq:inspiral}
{N}_{\mathrm{cycles}} = \frac{f}{2\pi} \frac{\mathrm{d}\phi}{\mathrm{d}f} = \frac{f^{2}}{\dot{f}} \; ,
\end{equation}
where an overdot represents the time derivative and $\dot{\phi} = 2\pi f$. The squared SNR scales with ${N}_{\mathrm{cycles}}$, so it would be expected that $h_\mathrm{c}(f)\approx \sqrt{{N}_{\mathrm{cycles}}}|\tilde{h}(f)|$.

The form for $h_\mathrm{c}$ can be derived from the Fourier transform in the stationary-phase approximation. Consider a source signal with approximately constant (root-mean-square) amplitude $h_0$ and central frequency $f'$. In this case,
\begin{eqnarray}
h(t) &=& \sqrt{2} h_0 \cos\left[\phi(t)\right] \; , \\
\tilde{h}(f) &=& \frac{h_0}{\sqrt{2}} \int_{-\infty}^{\infty}\rmd t\; \exp \left\{2\pi \rmi \left[\frac{\phi(t)}{2\pi t}-f\right]t\right\} + \exp \left\{-2\pi \rmi \left[\frac{\phi(t)}{2\pi t}+f\right]t\right\} \; .
\label{eq:fourier-h}
\end{eqnarray}
Without loss of generality, we can assume an initial phase of zero, such that $\phi(0) = 0$. The largest contribution to the integral comes from where the argument of the exponentials is approximately zero. For the first term, this occurs when $f = f'$, then the term in brackets is
\begin{equation}
\left[\frac{\phi(t)}{2\pi t}-f\right]_{f\,=\,f'} = f' + \dot{f'}t + \mathcal{O}\left(t^2\right) - f' = \dot{f'}t + \mathcal{O}\left(t^2\right).
\end{equation}
The higher-order terms cause the expoential to oscillate rapidly, such that these terms integrate to zero and may be neglected. Performing a similar expansion about $f = -f'$ for the second term in (\ref{eq:fourier-h}), and then evaluating the Gaussian integrals gives
\begin{eqnarray} \label{eq:FTofEMRI}
\tilde{h}(f) &\simeq& \frac{h_0}{\sqrt{2}} \int_{-\infty}^{\infty}\rmd t\; \exp\left(2\pi\rmi \dot{f'} t^{2}\right) + \exp\left(-2\pi\rmi \dot{f'} t^{2}\right) \nonumber \\
&\simeq& \frac{h_0}{\sqrt{2\dot{f'}}}.
\end{eqnarray}
From (\ref{eq:strain-hc}) and (\ref{eq:FTofEMRI}), the characteristic strain for inspiralling sources is given by \citep{FinnThorne}
\begin{equation}\label{eq:insphc}
h_\mathrm{c}(f) = \sqrt{\frac{2f^{2}}{\dot{f}}}h_0 \;.
\end{equation}
Equation (\ref{eq:strain-hc}) should be considered as the definition of characteristic strain and (\ref{eq:insphc}) a consequence of it for inspirals. Equation (\ref{eq:insphc}) is the relation between $h_\mathrm{c}(f)$ and the instantaneous root-mean-square amplitude $h_0$ for an inspiralling source; for other types of source a new relation satisfying (\ref{eq:hc}) has to be found.

\subsection{Stochastic backgrounds}\label{sec:stoch}

Another important source of GWs is that of stochastic backgrounds, which can be produced from a large population of unresolvable sources. These can be at cosmological distances, where it is necessary to distinguish the frequency in the source rest frame $f_{\mathrm{r}}$ from the measured frequency $f$; the two are related through the redshift $z$ via $f_{\mathrm{r}}=(1+z)f$. The comoving number density of sources $\nu$ producing the background is also a function of redshift; if the sources producing the background are all in the local Universe, then simply set $\nu(z) = \nu_0\delta (z)$ and replace $d_{\mathrm{L}}(z)$ with $d$ in all that follows, where $d_{\mathrm{L}}(z)$ and $d$ are respectively the luminosity and comoving distances to the source, $d_{\mathrm{L}}(z)=(1+z)d$.

We shall assume that the individual sources are binaries, in which case the number density of sources is also a function of the component masses. It is convenient to work in terms of the chirp mass, defined as ${\mathcal{M}}=\mu^{3/5}M^{2/5}$, where $\mu$ is the reduced mass and $M$ is the total mass of the binary. The comoving number density of sources shall be represented by $\nu(z, \mathcal{M})$.

Equation (\ref{eq:differentdescriptions}) gives an expression for the energy density in GWs per logarithmic frequency interval,
\begin{equation}\label{eq:stoch}
fS_{\mathrm{E}}(f)=\frac{\pi c^{2}}{4G}f^{2}\left[h_\mathrm{c}(f)\right]^2 \; .
\end{equation}
The total energy emitted in the logarithmic frequency interval $\mathrm{d}\left(\log f_{\mathrm{r}}\right)$ by a single binary in the population is $\left[\mathrm{d}E_{\mathrm{GW}}/\mathrm{d}(\log f_{\mathrm{r}})\right]\mathrm{d}(\log f_{\mathrm{r}})$; the energy density may be written as
\begin{equation}\label{eq:Phinney}
fS_{\mathrm{E}}(f) = \int_{0}^{\infty} \rmd z\; \frac{\mathrm{d}\nu}{\mathrm{d}z}\frac{1}{(1+z)}\frac{1}{\left[d_{\mathrm{L}}(z)\right]^2}\frac{\mathrm{d}E_{\mathrm{GW}}}{\mathrm{d}\left(\log f_{\mathrm{r}} \right)} \; , \end{equation}
where the factor of $\left( 1+z \right)^{-1}$ accounts for the redshifting of the energy.

For simplicity, consider the background to comprise of binaries in circular orbits, with frequencies $f_\mathrm{GW} =f_{\mathrm{r}}/2$, which are far from their last stable orbit. The energy radiated may then be calculated using the quadrupole approximation \citep{petersmathews1963}. The energy in GWs from a single binary per logarithmic frequency interval is
\begin{equation}\label{eq:Thorne}
\frac{\mathrm{d}E_{\mathrm{GW}}}{\mathrm{d}\left(\log f_{\mathrm{r}} \right)} = \frac{G^{2/3}\pi^{2/3}}{3}{\cal{M}}^{5/3}f_{\mathrm{r}}^{2/3}
\end{equation}
between minimum and maximum frequencies set by the initial and final radii of the binary orbit respectively. Here, we assume that the maximum and minimum frequencies are outside of the range of our detector and hence can be neglected. Using (\ref{eq:stoch}), (\ref{eq:Phinney}) and (\ref{eq:Thorne}), an expression for characteristic strain can now be found \citep{SesanaVecchioColancino}
\begin{equation}\label{eq:bigint}
\left[h_\mathrm{c}(f)\right]^2 = \frac{4G^{5/3}}{3\pi^{1/3}c^{2}}f^{-4/3}\int_{0}^{\infty}\mathrm{d}z\;\int_{0}^{\infty}\mathrm{d}{\cal{M}}\;\frac{\mathrm{d}^{2}\nu}{\mathrm{d}z\,\mathrm{d}{\mathcal{M}}}\frac{1}{\left[d_{\mathrm{L}}(z)\right]^2}\left( \frac{{\mathcal{M}}^{5}}{1+z} \right)^{1/3}\; .
\end{equation}

From (\ref{eq:bigint}) it can be seen that the characteristic strain due to a stochastic background of binaries is a power law in frequency with spectral index $\alpha=-2/3$. The amplitude of the background depends on the population statistics of the binaries under consideration via $\nu(z,{\mathcal{M}})$. The power law is often parametrised as
\begin{equation}\label{eq:power} 
h_\mathrm{c}(f) = A\left(\frac{f}{f_{0}}\right)^{\alpha}\; , 
\end{equation}
and constraints are then placed on $A$. In practice, this power law also has upper and lower frequency cut-offs related to the population of source objects.

A stochastic background from other sources, such as cosmic strings or relic GWs from the early Universe, can also be written in the same form as (\ref{eq:power}), but with different spectral indices: $\alpha=-7/6$ for cosmic strings or $\alpha$ in the range $-1$ to $-0.8$ for relic GWs \citep{Jenet}.

An alternative method for graphically representing the sensitivity of a GW detector to stochastic backgrounds, called the \emph{power-law-integrated sensitivity curve}, was suggested by \cite{2013PhRvD..88l4032T}. This method accounts for the there being power across all frequencies in the sensitivity band by integrating the noise-weighted signal over frequency. As our aim here is to present stochastic backgrounds alongside other types of sources for comparison, we will not use this approach.

\subsection{Burst sources}\label{sec:bursts}

Some sources of GWs can produce signals with large amplitudes, greater than the detector noise. The typical duration of such a signal is short, of the order of a few wave periods, and so there is not time to accumulate SNR in each frequency band as for inspirals. As a consequence, waveform models are not required for detection; we simply rely on identifying the excess power produced by these burst sources. Typically, we may be looking for signals from core-collapse supernovae \citep{Ott2009}, the late stages of merging compact binaries \citep{Abadie2012}, cosmic strings \citep{Aasi2014}, or more generally, signals from any unexpected or poorly modelled sources.

Burst searches are often carried out using time--frequency techniques. The data stream from a detector is temporally split into segments, the length of which can be tuned to give greater sensitivity to particular sources. Each segment is then transformed into the frequency domain, whitened and normalised to the noise spectrum of the detector to produce a time--frequency plot. Potential GW signals are identified by searching for clusters of pixels that contain an excess of power \citep[e.g.,][]{Bursts}.

The presence of excess power across a number of pixels eliminates modelled noise sources, but such a cluster may also be caused by atypical noise within a detector. We can improve our confidence of a GW signal by making use of information obtained from other GW detectors. Signals across a network of detectors should have compatible arrival times (given the sky direction) as well as consistent amplitudes, frequencies and shapes of the waveform. Different pipelines are currently in use that analyse the signal consistency in different ways: both coincidence searches \citep{Chatterji2004} and fully coherent methods \citep{Klimenko2008} are used.

An important aspect of burst search algorithms is to accurately estimate the noise properties within each time segment. To this extent, null data streams can be constructed that are insensitive to real GW signals. In order to estimate the false alarm rate, the data from different detectors can be shifted in time to remove any genuine coincident GW signals \citep{Cannon2008}. These time-shifts are then analysed to simulate the potential occurence of coincident noise events. The algorithms are tuned using time-shifted data to ensure there is no bias in the final search.

As discussed in \ref{sec:insp}, the expected relation between $h_\mathrm{c}(f)$ and a typical waveform $\tilde{h}(f)$ is
\begin{equation}\label{eq:simple} 
h_\mathrm{c}(f) = \sqrt{{N}_{\mathrm{cycles}}}\left|\tilde{h}(f)\right| \; , 
\end{equation}
where ${N}_{\mathrm{cycles}}$ is the number of cycles of radiation generated by the source, which is of order unity for bursts. 

An alternative characterisation of the signal amplitude commonly used for burst sources is the root-sum-square of the waveform polarisations:
\begin{equation}
h_\mathrm{rss}^2 = \int \rmd t \; {|h_+(t)|}^2 + {|h_\times(t)|}^2 \; .
\end{equation}
For a linearly polarised GW, with $\tilde{h}(f)$ constant across the bandwidth $\Delta f$, this is approximately related to the characteristic strain via 
\begin{equation}
h_\mathrm{rss} \simeq \left|\tilde{h}(f)\right|\sqrt{\Delta f}\;,\label{eq:rela}
\end{equation}
where we have neglected the detector response functions (see section \ref{sec:principles}), which are of order unity. In this work, we favour a constant $h_\mathrm{c}(f)$ rather than $h_\mathrm{rss}$ for consistency with the other types of source where the bandwidth is detector specific.

\section{Detectors}\label{sec:detectors}

In this section we introduce the detector noise curves used in \ref{app:a}. We begin with a description of the basic operation of detectors. We then discuss ground-based detectors, space-based detectors and PTAs in turn. The latter function somewhat differently than conventional interferometers, so we include a brief introduction to PTA analysis. References for the noise curves used for individual detectors can be found in the relevant subsections and further information about the detectors can be discovered in these. Detectors are frequently upgraded and redesigned, hence, while these curves are believed to be correct at the time of writing, it is best to check for updates from the appropriate science teams before relying on the details given here, although we hope that they shall remain accurate enough for illustrative purposes.

\subsection{Operating principle of an interferometric detector}\label{sec:principles}

All of the man-made detectors discussed in this section utilise the principle of interferometry. Such detectors work by taking a beam of monochromatic light and splitting it into two beams travelling at some angle to each other. Each beam is passed in to an optical cavity where it undergoes a number of round trips before being recombined to form an interference pattern. The ends of the cavity are, in the ideal case, freely floating test masses which move relative to each other in response to a passing GW, this effect is measured by observing the changing interference pattern.

The response of a detector to an incident plane-fronted GW depends upon the relative orientations of the detector and the incoming wave. Let us choose the origin of our coordinate system to be the beam-splitter of the interferometer, and $l_{1}^{i}$ and $l_{2}^{i}$ to be unit 3-vectors pointing along the two arms. In the absence of noise the output of the detector is the difference in strain between the two arms \citep{Thorne1987}
\begin{equation}\label{eq:hoftxx}
h(t)=\frac{1}{2}h_{ij}\left( l_{1}^{i}l_{1}^{j}-l_{2}^{i}l_{2}^{j} \right)\; ,
\end{equation}
where $h_{ij}$ are the spatial components of the GW metric perturbation. Let $\hat{r}^{i}$ be the unit 3-vector pointing towards the source of the GWs, with spherical polar angles $(\theta,\phi)$ relative to some axes fixed to the detector, and let $p^{i}$ and $q^{i}$ be unit vectors orthogonal to $\hat{r}^{i}$. We can now define the basis tensors
\begin{eqnarray}
H^{+}_{ij}&=p_{i}p_{j}-q_{i}q_{j} \; , \\
H^{\times}_{ij}&=p_{i}q_{j}+q_{i}p_{j} \; .
\end{eqnarray}
There remains a freedom in the coordinates described, a rotation of $p^{i}$ and $q^{i}$ through an angle $\psi$ about $\hat{r}^{i}$ known as the polarization angle. For a single frequency component, the strain induced by a GW may be written as
\begin{equation}\label{eq:hijxx}
h_{ij}=A_{+}H^{+}_{ij}\cos\left(2\pi ft\right)+A_{\times}H^{\times}_{ij}\cos\left(2\pi ft+\Delta \phi\right) \; ,
\end{equation}
where $A_{+}$ and $A_{\times}$ are the amplitudes of the two polarisation states. Combining (\ref{eq:hoftxx}) and (\ref{eq:hijxx}) allows the detector output to be written as
\begin{equation}
h(t)=F^{+}(\theta,\phi,\psi)A_{+}\cos\left(2\pi ft\right)+F^{\times}(\theta,\phi,\psi)A_{\times}\cos\left(2\pi f t + \Delta\phi \right)\; ,
\end{equation}
where the response functions inherit their angular dependence from the choice of coordinates
\begin{eqnarray}\label{eq:responsefuncs}
F^{+}(\theta,\phi,\psi)&=\frac{1}{2}H^{+}_{ij}\left(l_{1}^{i}l_{1}^{j}-l_{2}^{i}l_{2}^{j}\right) \; , \\
F^{\times}(\theta,\phi,\psi)&=\frac{1}{2}H^{\times}_{ij}\left(l_{1}^{i}l_{1}^{j}-l_{2}^{i}l_{2}^{j}\right) \; . 
\end{eqnarray}

\begin{figure}
 \centering
 \includegraphics[trim=0cm 0cm 0cm 0cm, width=0.99\textwidth]{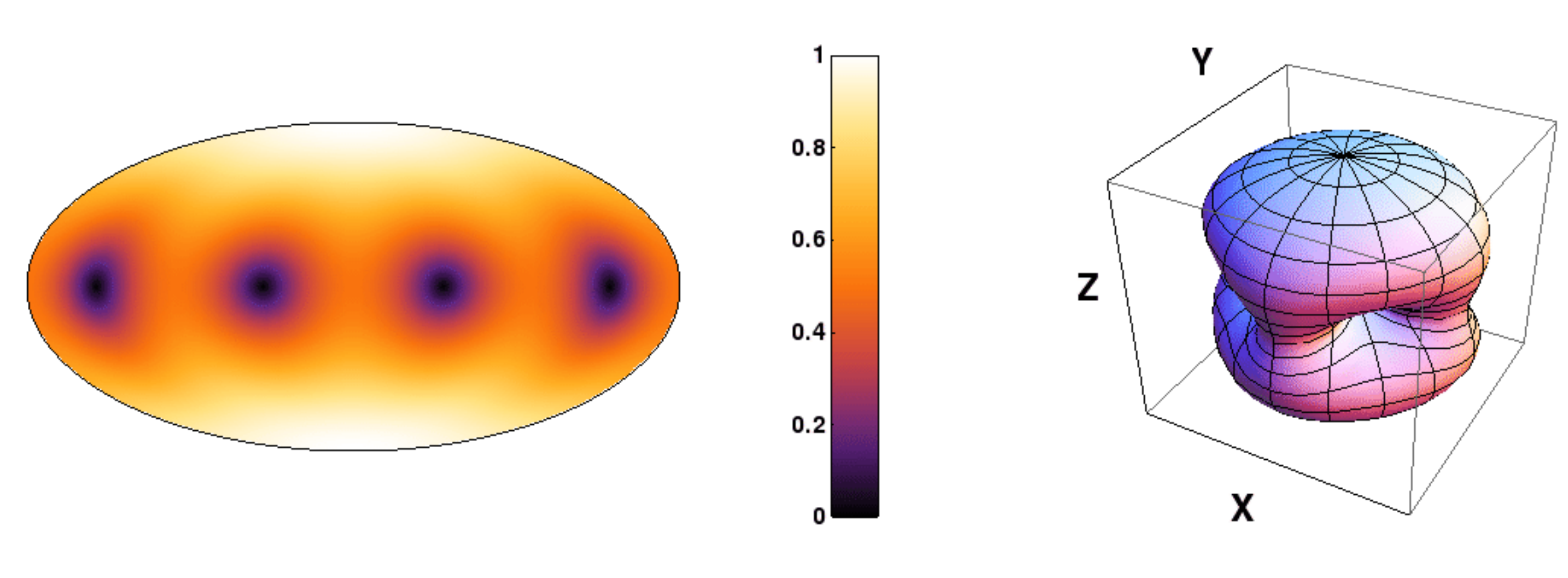}
 \caption{The angular response function of an interferometric detector shown both as a surface plot and in an Aitoff--Hammer projection. The quantity that is plotted is is the polarisation average, $\left[(1/2\pi)\int\mathrm{d}\psi \;(F^{+})^{2}+(F^{\times})^{2}\right]^{1/2}$. The response is a function of two sky angles, $\theta$ and $\phi$, and varies between $0$ and $1$. The two detector arms lie in the $x$--$y$ plane either side of one of the zeros in the response.}
 \label{fig:LIGO}
\end{figure}

The response function of a two-arm interferometric detector is quadrupolar, an example is plotted in figure \ref{fig:LIGO}.
Throughout this paper, detector sensitivity refers to the polarisation and sky averaged sensitivity $F$, where
\begin{equation}\label{eq:skyav}
F^{2}=\int_{0}^{2\pi}\frac{\mathrm{d}\psi}{2\pi}\; \int_{0}^{2\pi} \frac{\mathrm{d}\phi}{2\pi}\; \int_{0}^{\pi}\frac{\sin\theta\,\mathrm{d}\theta}{2}\;\left[F^{+}\left(\theta,\phi,\psi\right)^{2}+F^{\times}\left(\theta,\phi,\psi\right)^{2}\right]\; .
\end{equation}
For a single $90\degree$-interferometer, such as LIGO, the sky and polarisation averaged response is $F=\sqrt{1/5}\approx 0.447$.

A detector may consist of several interferometers. Let $F_{a}$ be the averaged response of the $a$-th interferometer, the average response of a network of $k$ detectors is obtained by adding in quadrature,
\begin{equation} F_{\mathrm{Total}}^{2}=\frac{1}{k}\sum_{a\,=\,1}^{k}F_{a}^{2} \; .\end{equation}

The averaging in (\ref{eq:skyav}) assumes a uniform distribution of polarisation angles $\psi$. This is the case for a stochastic background; however, for a non-inspiralling circular binary, the polarisation is a function of the two spherical polar angles ($\iota,\xi$) specifying the orientation of the binary's orbital angular momentum. Here, $\iota$ is the inclination angle, the polar angle between the orbital angular momentum and the line joining the source to the detector ($-\hat{r}$) and $\xi$ is the azimuthal angle around the same line. In this case, to characterise the detector sensitivity we average over all four angles ($\theta,\phi,\iota,\xi$). If the binary is inspiralling, then the polarisation depends on still more parameters which need to be averaged over. These more complicated averages all have the property that they depend on both the detector and the source, hence they are unhelpful for our present purpose separating the source amplitude from the detector sensitivity. Additionally, the different averages do not work out to be so different from each other: \citet{1993PhRvD..47.2198F} calculated the sensitivity for a detector with the LIGO geometry averaged over the four angles ($\theta,\phi,\iota,\xi$) as $\sqrt{4/25}=0.4$ times peak sensitivity, which should be compared with the value $\sqrt{1/5}\approx 0.447$ above. The effect of replacing the true sensitivity with the sky-averaged sensitivity for the LISA detector was considered in detail by \cite{2012CQGra..29l4015V}; they also found there is a small difference when considering an entire population of sources. For the remainder of this paper the three-angle average defined in (\ref{eq:skyav}) is used.

\subsection{Ground-based detectors}\label{sec:ground}

Ground-based detectors are the most numerous. A collection of interferometric detectors are listed in table \ref{table:t}, these are sensitive to GWs in the frequency range ${\mathcal{O}}(10$--$10^{3})~\mathrm{Hz}$. They all simulate free-floating test masses by suspending a mass from a pendulum system with natural frequency far removed from than that of the GW. Their sensitivity curves include narrow lines that arise from noise sources in the instrument, including resonances in the suspension system and electrical noise at multiples of $60~\mathrm{Hz}$: these have been removed in the \ref{app:a} figures for clarity. The detectors fall broadly into three categories: first-generation detectors, which have already operated; second-generation detectors currently under construction; and third-generation detectors at the planning stage. 

The most notable ground-based detectors are LIGO and Virgo, which work in collaboration, supported by GEO600. LIGO, Virgo and GEO600 have completed science runs as first-generation detectors \citep[e.g.,][]{Abadie2010}. As LIGO and Virgo are currently being upgraded their initial configurations are now referred to as Initial LIGO (iLIGO) and Initial Virgo (iVirgo) respectively. The upgraded, second-generation versions are referred to as Advanced LIGO (aLIGO) and Advanced Virgo (AdV) respectively. LIGO has two observatories: one at Hanford, Washington, which has two detectors; and another at Livingston, Louisiana. There is an agreement to move one of the upgraded Hanford detector systems to a location in India \citep{LIGO-India,Unnikrishnan2013}. The GEO600 detector is not subject to a major upgrade plan, but since summer 2009 it has been enhanced by a series of smaller improvements, notably improving high-frequency sensitivity \citep[GEO-HF;][]{2006CQGra..23S.207W}. The advanced detectors should start operation in the next couple of years, with LIGO-India following further in the future.

TAMA300 is a Japanese first-generation detector. Its successor, currently under construction, is the Kamioka Gravitational Wave Detector (KAGRA), formerly the Large-scale Cryogenic Gravitational wave Telescope (LCGT), which is located underground in the Kamioka mine. It employs more sophisticated noise-reduction techniques than LIGO or Virgo, such as cryogenic cooling.

The Einstein Telescope (ET) is an ambitious proposal to construct an underground third-generation detector. Its location would provide shielding from seismic noise, allowing it to observe frequencies of $(10$--$10^4)~\mathrm{Hz}$. 

\begin{table}
\caption{\label{table:t} Summary of ground-based laser interferometers.}
\begin{indented}
\item[]\begin{tabular}{ l l l l l }
\br
{\bf Detector} & {\bf Country} & {\bf Arm length} & {\bf  Approximate date} & {\bf Generation} \\
\mr
  GEO600$^{a}$	&	Germany 	& $600~\mathrm{m}$ 	& 2001--present	   & First \\
  TAMA300$^{b}$ & 	Japan		& $300~\mathrm{m}$ 	& 1995--present    & First \\
  iLIGO$^{c}$	&	USA		& $4~\mathrm{km}$ 	& 2004--2010 	   & First \\
  Virgo$^{d}$	& 	Italy		& $3~\mathrm{km}$ 	& 2007--2010 	   & First \\
  aLIGO$^{e}$	&	USA		& $4~\mathrm{km}$ 	& \emph{est.} 2015 & Second \\
  AdV$^{f}$	&	Italy	 	& $3~\mathrm{km}$ 	& \emph{est.} 2016 & Second \\
  KAGRA$^{g}$	&	Japan		& $3~\mathrm{km}$ 	& \emph{est.} 2018 & Second \\
  ET$^{h}$      & 	---		& $10~\mathrm{km}$ 	& \emph{est.} 2025 & Third \\
\br
\multicolumn{5}{l}{$^{a}$\citet{2010CQGra..27h4003G}, $^{b}$\citet{2002CQGra..19.1409A}, $^{c}$\citet{2009RPPh...72g6901A},}\\
\multicolumn{5}{l}{$^{d}$\citet{accadia_virgo:2012}, $^{e}$\citet{2010CQGra..27h4006H}, $^{f}$\citet{Acernese2009}, $^{g}$\citet{2012CQGra..29l4007S}, $^{h}$\citet{2011CQGra..28i4013H}.}\\
\br
\end{tabular}
\end{indented}
\end{table}

We use an interpolation to the data published on \url{https://wwwcascina.virgo.infn.it/advirgo/} (2013) for the AdV sensitivity curve, an interpolation to the data for version D of the KAGRA detector published on \url{http://gwcenter.icrr.u-tokyo.ac.jp/en/researcher/parameter} (2013) and analytic fits to the sensitivity curves from \citet{Sathyaprakash} for the remaining detectors.

\subsection{Space-based detectors}\label{sec:space}

Space-based detectors work on similar principles to ground based detectors, but with the test masses residing inside of independent, widely separated satellites. Space-based detectors are sensitive to lower frequency GWs than their ground-based counterparts; this is partly because space-based detectors can have much longer arms, and partly because they are unaffected by seismic noise which limits the low frequency performance of ground-based detectors. 

The canonical design for a space-based detector is the Laser Interferometer Space Antenna (LISA), which is sensitive to millihertz GWs. LISA would consist of three satellites flying in a triangular constellation with arms of length $5\times 10^{9}~\mathrm{m}$ in a $1~\mathrm{AU}$ orbit around the Sun, trailing the Earth by $20\degree$. The laser arms in a LISA-like detector are not a cavity, the light only travels once along each arm. eLISA is a rescoped version of LISA designed to probe the same frequency range, while proposals such as the Advanced Laser Interferometer Antenna (ALIA), Big Bang Observer (BBO) and Deci-hertz Interferometer GW Observatory (DECIGO) are designed to probe decihertz GWs.

\subsubsection{LISA and eLISA}

The instrumental noise curves for LISA are approximated by the analytic fit given by \citet{Sathyaprakash}, which we use for the plots in \ref{app:a}. When observing individual sources with LISA there is an additional contribution to the noise from a background of unresolvable binaries. This is not included here as we consider the background as a source of GWs (see section \ref{sec:GB}). eLISA is a rescoped version of the classic LISA mission, the main differences are shorter arms ($10^{9}~\mathrm{m}$ instead of $5\times 10^{9}~\mathrm{m}$), two laser arms instead of three, and a different orbit (drifting away from Earth instead of $20\degree$ Earth trailing). The effect of these changes is a slightly reduced peak sensitivity and a shift to higher frequencies. We use an analytic fit to the instrumental noise curve given by \citet{Amaro-Seoane-et-al}.

\subsubsection{DECIGO, ALIA and BBO}

These missions are designed to probe the decihertz region of the GW spectrum; they are considerably more ambitious than the LISA or eLISA mission and their launches will be further into the future. We use a simple analytic fit to the sensitivity curve for ALIA~\citep{bender_possible_2013}, while for DECIGO and BBO, fits to the sensitivity curves given by \citet{2011PhRvD..83d4011Y} are used.

\subsection{Pulsar timing arrays}\label{sec:PTAgeneralproperties}

PTAs can be thought of as naturally occurring interferometers with galactic-scale arm lengths. Accordingly, they are sensitive to much lower frequencies than the detectors previously discussed. Each pulsar is a regular clock and the measured pulse arrival time can be compared against a prediction, leaving a residual which includes the effects of passing GWs. Using an array of these pulsars spread across the sky allows us to correlate residuals between different pulsars, to exploit the fact that GWs influence all pulsars whereas intrinsic pulsar noise does not. The correlation between different pulsars depends only on their angular separation on the sky, and has a distinctive shape, known as the Hellings and Downs curve \citep{HellingsDowns}.

\begin{figure}
 \centering
 \includegraphics[trim=0cm 0cm 0cm 0cm, width=0.55\textwidth]{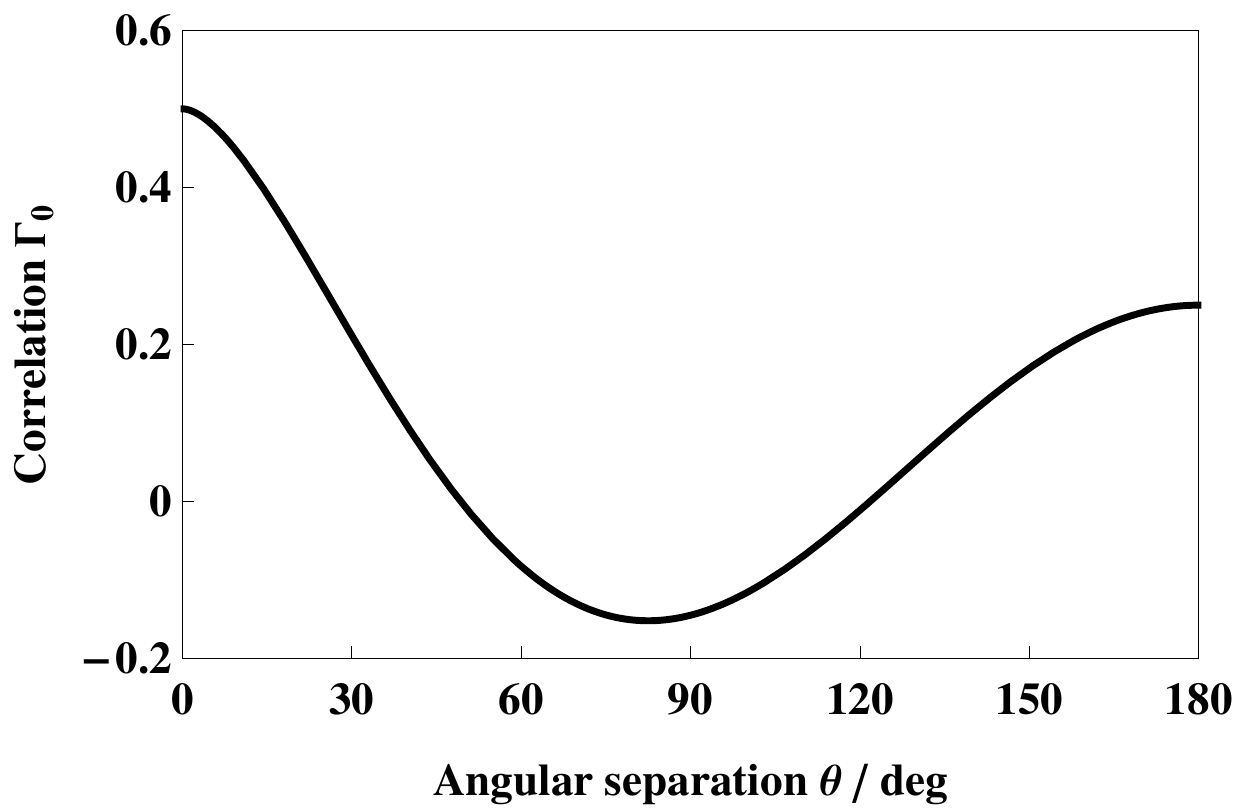}
 \caption{The \citet{HellingsDowns} curve, the correlation between two pulsars separated on the sky by an angle $\theta$.}
 \label{fig:HnD}
\end{figure}

The redshift of the rate of arrival of pulses for a pulsar at a distance $L$ from the Solar-System barycentre (SSB), in the direction of the unit spatial vector $\hat{p}$ induced by a GW travelling in direction of the unit vector $\hat{\Omega}$ is \citep{anholm-2009}
\begin{equation}\label{eq:pulsarredshift}
z(t,\hat{\Omega})= \frac{1}{2} \frac{\hat{p}^{j}\hat{p}^{i}} {1+\hat{\Omega}\cdot\hat{p}}\left[h_{ij}^{\mathrm{Pulsar}}\left(t-\frac{L}{c},\hat{\Omega} \right)-h_{ij}^{\mathrm{Earth}}(t,\hat{\Omega} )\right] = \frac{1}{2}\frac{\hat{p}^{j}\hat{p}^{i}}{1+\hat{\Omega}\cdot\hat{p}} \Delta h_{ij}(t,\hat{\Omega})\; .
\end{equation}
The redshift includes two terms: the pulsar term and the Earth term. The pulsar term is often neglected in PTA analysis as it can be considered as an extra noise term which averages to zero across the array. The experimentally measured quantity is not the redshift but the timing residual, the two are related via
\begin{equation}\label{eq:restored}
R(t,\hat{\Omega}) = \int_{0}^{t}\mathrm{d}t'\;z(t',\hat{\Omega}) \; .
\end{equation}
All of the pulsars, and the Earth, are subject to the same metric-perturbation field. This may be expressed in terms of its Fourier transform
\begin{equation}
h_{ij}(t,\vec{r}) = \sum_{A\,=\,+,\,\times}\int\mathrm{d}f\;\iint_{\mathbb{S}_{2}}\mathrm{d}\hat{\Omega}\; \tilde{h}_{A}(f,\hat{\Omega})e_{ij}^{A}(\hat{\Omega}) \exp\left[2\pi i f \left(t-\frac{\hat{\Omega}\cdot\vec{x}}{c}\right)\right] \; ,
\end{equation}
where $e^{A}_{ij}(\hat{\Omega})$ is the $A$ polarisation basis tensor for direction $\hat{\Omega}$, and $\vec{r}$ is the spatial position. Choosing the SSB as the origin of our coordinate system, so the pulsar is at position $L\hat{p}$, gives
\begin{equation}\label{eq:deltahij}
\Delta h_{ij}(t,\hat{\Omega}) = \sum_{A\,=\,+,\,\times}\int\mathrm{d}f\; \tilde{h}_{A}(f,\hat{\Omega})e_{ij}^{A}(\hat{\Omega})\exp\left(2\pi \rmi f t\right)\left\{\exp\left[-2\pi\rmi fL \left(1+\hat{p}\cdot\hat{\Omega}\right)\right]-1\right\} \; .
\end{equation}
From (\ref{eq:pulsarredshift}) and (\ref{eq:deltahij}), the Fourier transform of the redshift $\tilde{z}(f,\hat{\Omega})$ can be identified as
\begin{equation}\label{eq:45}
\tilde{z}(f,\hat{\Omega}) = \left\{\exp\left[-2\pi\rmi fL \left(1+\hat{p}\cdot\hat{\Omega}\right)\right]-1\right\} \sum_{A\,=\,+,\,\times} \tilde{h}_{A}(f,\hat{\Omega})F^{A}(\hat{\Omega}) \; ,
\end{equation}
where
\begin{equation}
F^{A}(\hat{\Omega}) = \frac{e_{ij}^{A}(\hat{\Omega})\hat{p}^{j}\hat{p}^{i}} {2\left(1+\hat{\Omega}\cdot\hat{p}\right)} \, .
\end{equation}
The function $F^{A}(\hat{\Omega})$ may be regarded as the PTA equivalent of the detector response functions in (\ref{eq:responsefuncs}). The stochastic background of GWs is fully characterised by the one-sided PSD via the expectation value
\begin{equation}\label{eq:stochback}
\left<\tilde{h}_{A}^{*}(f,\hat{\Omega})\tilde{h}_{A'}(f',\hat{\Omega}')\right> = \frac{1}{2}S_{h}(f)\delta^{(2)}(\hat{\Omega}, \hat{\Omega}')\delta_{AA'}\delta(f-f') \; ,
\end{equation}
where $\delta^{(2)}(\hat{\Omega},\hat{\Omega}')$ is the delta-function on the sphere. From (\ref{eq:45}) and (\ref{eq:stochback}), the expectation of the product of signals from two different pulsars in directions $\hat{p}_{1}$ and $\hat{p}_{2}$ may be evaluated as
\begin{equation} 
\left<\tilde{z}_{1}(f)\tilde{z}_{2}^{*}(f')\right> = \frac{1}{2}S_{h}(f)\delta (f-f')\Gamma(f) \; ,
\end{equation}
where
\begin{eqnarray}
\Gamma(f) &= \sum_{A\,=\,+,\,\times} & \iint_{{\mathbb{S}}^{2}}\mathrm{d}\hat{\Omega}\,\left\{\exp\left[2\pi\rmi fL_{1}\left(1+\hat{\Omega}\cdot\hat{p}_{1}\right)\right]-1\right\} \nonumber \\*
 & & \times \left\{\exp\left[-2\pi\rmi fL_{2}\left(1+\hat{\Omega}\cdot\hat{p}_{2}\right)\right]-1\right\} F_{1}^{A}(\hat{\Omega})F_{2}^{A}(\hat{\Omega}) \; .
\end{eqnarray}
The overlap function $\Gamma(f)$ tends to a constant value in the limit that the distances to the pulsars are large compared to the wavelength of GWs; PTAs operate in this limit \citep{Mingarelli2014}, so the overlap may be approximated as a constant, 
\begin{equation}
\Gamma(f) \approx \Gamma_{0} = \sum_{A\,=\,+,\,\times}\iint_{{\mathbb{S}}^{2}}\mathrm{d}\hat{\Omega}\;F_{1}^{A}(\hat{\Omega})F_{2}^{A}(\hat{\Omega})\;.
\end{equation}
Neglecting the exponential terms in the overlap is the frequency-domain equivalent of neglecting the pulsar term in (\ref{eq:pulsarredshift}). The integral may be evaluated to give an expression depending only on the angle $\theta$ between the two pulsars; this is the famous Hellings and Downs curve, shown in figure \ref{fig:HnD},
\begin{equation}
\Gamma_{0} = \frac{1}{2}+\frac{3\xi}{2}\left(\ln \xi -\frac{1}{6}\right) \; ,
\end{equation}
where $\xi = (1-\cos\theta)/{2}.$

The sensitivity bandwidth of a PTA is set by the sampling properties of the data set. If measurements are spaced in time by $\delta t$ and taken for a total length of time $T$, then the PTA is sensitive to frequencies in the range $(1/T) < f < (1/\delta t)$. The characteristic strain that the PTA is sensitive to scales linearly with $f$ in this range. This gives the wedge-shaped curves plotted in \ref{app:a}. The absolute value of the sensitivity is fixed by normalising to a calculated limit at a given frequency for each PTA. For a discussion of the sensitivities of PTA to both individual sources and stochastic background, see \citet{MooreTaylorGair}.

There is a discrepancy between the treatment of PTA sensitivity curves here and the higher frequency detectors discussed in sections \ref{sec:ground} and \ref{sec:space}. When observing a long-lived source, such as an inspiral, with a high frequency detector, the convention was to define a \emph{characteristic} strain to satisfy (\ref{eq:hc}). Here, the convention is to leave the strain untouched and instead adjust the PTA sensitivity curve with observation time, again to satisfy (\ref{eq:hc}). This discrepancy is an unfortunate result of the conventions in use by the different GW communities; however, it is natural given the sources under observation. When observing a transient source, such as a burst or inspiral, which changes within the lifetime of the detector, it is natural to consider the detector as performing constantly while the signal changes. However, when observing a monochromatic source or a stochastic background, which is unchanging over the detector lifetime, it is more natural to consider the source as being fixed and the sensitivity of the detector gradually improving. All that is required by the definition in (\ref{eq:hc}) is that the ratio $h_\mathrm{c}(f)/h_{n}(f)$ is constant.

\subsubsection{Current PTAs}

The PTAs currently in operation are the European Pulsar Timing Array \citep[EPTA\footnote{\url{http://www.epta.eu.org/}};][]{eptareview2013}, the Parkes Pulsar Timing Array \citep[PPTA\footnote{\url{http://www.atnf.csiro.au/research/pulsar/ppta/}};][]{parkesreview2013} based in Australia, and the North American Nanohertz Observatory for Gravitational waves \citep[NANOGrav\footnote{\url{http://nanograv.org/}};][]{nanogravreview2013}. There are published limits on the amplitude of the stochastic background from all three detectors: the most recent from EPTA is \citet{Haasteren}, from PPTA is \citet{Shannon2013} and from NANOGrav is \citet{2013ApJ...762...94D}. The limits from the existing detectors are all of a similar magnitude.

In \ref{app:a} we use the EPTA limit based on an analysis of $5$ pulsars over approximately $10~\mathrm{years}$. The curve labelled ``EPTA'' assumes that all the pulsar were timed identically every 2 weeks with an r.m.s.\ error in each timing residual of $100~\textrm{ns}$. The NANOGrav limit is comparable to this, and the PPTA limit is a factor $\sim 2.5$ lower.

\subsubsection{The International Pulsar Timing Array}

Combining the existing arrays would yield a single PTA using approximately $4$ times as many pulsars; this consortium of consortia is known as the International Pulsar Timing Array \citep[IPTA\footnote{\url{http://www.ipta4gw.org/}};][]{iptareview2013}. The IPTA curve plotted in the figures assumes $20$ pulsars timed every 2 weeks for 15 years with an r.m.s.\ error in each timing residual of $100~\textrm{ns}$.

\subsubsection{SKA}

The next great advancement in radio astronomy shall come with the completion of the Square Kilometre Array \citep[SKA;][]{Dewdney2009}. This shall greatly increase the sensitivity of pulsar timing \citep{Kramer2004}. The sensitivity curves plotted in \ref{app:a} for the SKA assumes $50$ pulsars timed every 2 weeks for 20 years with an r.m.s.\ error in each timing residual of $30~\textrm{ns}$. Our choice of 50 pulsars might be conservative, it is possible that the SKA will discover many more suitable milli-second pulsars than this throughout its operation, in which case the corresponding sensitivity curve would become lower.

\section{Astrophysical sources}\label{sec:sources}

All the sources described here are represented by shaded boxes in \ref{app:a}. Sources with short durations (i.e., burst sources) and sources that evolve in time over much longer timescales than our observations are drawn with flat-topped boxes for $h_\mathrm{c}(f)$. Inspiraling binaries, or stochastic backgrounds of binaries, are drawn with a sloping top proportional to $f^{-2/3}$ for $h_\mathrm{c}(f)$, which is result derived in section \ref{sec:stoch}. The width of the box gives the range of frequencies sources of a given type can have while remaining at a detectable amplitude. The question of the height of the box is more problematic; it would be desirable to normalise each box so that there was a fixed event rate, say one event per year with an amplitude lying within the box. However, for many of the sources considered, the event rate is subject to a large degree of uncertainty (and estimates change rapidly as our understanding of the astrophysics improves). Consequently, we take the more definite, but somewhat arbitrary, approach that for each type of source a fiducial event is nominated (with parameters or amplitude detailed in the relevant section below) and the amplitude of this event is used to position the top of the box. The parameters for each fiducial event are chosen such that the resulting amplitude is roughly consistent with the optimistic end of current predictions: the boxes indicate where we \emph{could} plausibly find GW sources; the actual event rates could turn out to be substantially lower than this upper bound. 

\subsection{Sources for ground-based detectors}

\subsubsection{Neutron-star binaries}

The inspiral and merger of a pair of neutron stars is the primary target for ground-based detectors. The expected event rate for this type of source is uncertain, but estimates centre around $\gamma_{\mathrm{NS-NS}}=1.3\times 10^{-4}~\mathrm{Mpc^{-3}\,yr^{-1}}$ \citep{CBC}. Plotted in \ref{app:a} are boxes labelled ``\emph{compact binary inspirals}'' with amplitudes such that a ratio $h_\mathrm{c}/h_{n}=16$ is produced for Advanced LIGO at peak sensitivity and a width between $(3$--$300)~\mathrm{Hz}$, corresponding to expected observable frequencies.

\subsubsection{Supernovae}

Simulations of core-collapse supernovae show that GWs between $(10^{2}$--$10^{3})~\mathrm{Hz}$ can be produced \citep{Kotake2006}. The GW signal undergoes ${\mathcal{O}}(1)$ oscillation and is hence burst-like. \citet{2002A&A...393..523D} calculate the average maximum amplitude of GWs for a supernova at distance $r$ as
\begin{equation}
h_\mathrm{max} = 8.9\times 10^{-21}\left( \frac{10~\mathrm{kpc}}{r} \right) \; .
\end{equation}
The event rate for supernovae is approximately $\gamma_{\mathrm{SN}} = 5\times10^{-4}~\mathrm{Mpc^{-3}\,yr^{-1}}$. The boxes labelled ``\emph{supernova}'' plotted in \ref{app:a} correspond to a distance $r=300~\mathrm{kpc}$ with the frequency range quoted above. The LIGO and Virgo detectors have already placed bounds on the event rate for these sources \citep{Bursts}.

\subsubsection{Continuous waves from rotating neutron stars}

Rotating neutron stars are a source of continuous GWs if they possess some degree of axial asymmetry \citep{Abbott2007, Prix2009, Einstein@Home}. The signals are near monochromatic with a frequency twice the rotation frequency of the neutron star, and are a potential source for ground-based detectors. The amplitude of the GWs depends upon the deformation of the neutron star, and its spin frequency. The magnitude of the distortion depends upon the neutron star equation of state and the stiffness of the crust, which are currently uncertain \citep{Chamel2008,Lattimer2012}. Deformations can also be supported by internal magnetic fields \citep{Haskell2008}. Several known pulsars could be sources for the advanced detectors and upper limits from the initial detectors help to constrain the deformations. 

The boxes labelled ``\emph{pulsars}'' plotted in \ref{app:a} correspond to the upper limits placed on a GW signal from the Crab pulsar \citep{Aasi2014a}, extrapolated across a frequency range between $(20$--$10^{3})~\mathrm{Hz}$. The extrapolation was performed using the scaling $h_0 \propto \sqrt{\dot{f}/f}$~\citep{Aasi2014a} (this is the spin-down limit, which assumes that the only loss of energy from the system is due to GW emission) and (\ref{eq:insphc}). This gives $h_\mathrm{c} \propto f^{1/2}$; higher frequency sources are observed at a louder SNR because they undergo more cycles.

\subsection{Sources for space-based detectors}

For a review of the GW sources for space-based missions see, for example, \citet{Amaro-Seoane-et-al}, \citet{Gairetal} or \citet{eLISAyellowbook}.

\subsubsection{Massive black-hole binaries}

Space-based detectors shall be sensitive to equal-mass mergers in the range $(10^{4}$--$10^{7})\,\Msun$. Predictions of the event rate for these mergers range from ${\mathcal{O}}(10$--$100)~\mathrm{yr}^{-1}$ for eLISA with SNRs of up to $10^3$ \citep{TheGravitationalUniverse}. The large range in the rate reflects our uncertainty in the growth mechanisms of the supermassive black hole population \citep{Volonteri2010}. A $10^6 \Msun$ binary can be observed out to a redshift $z \sim 8$ with an SNR of $100$ \citep{TheGravitationalUniverse}. This fiducial source gives the amplitude of the boxes labelled ``\emph{$\mathit{\approx 10^{6}}$ solar mass binaries}'' in \ref{app:a}. The range of frequencies is $(3\times 10^{-4}$--$3\times 10^{-1})~\mathrm{Hz}$, extrapolated using the slope $h_\mathrm{c}(f) \propto f^{-2/3}$ derived in section \ref{sec:stoch}.

\subsubsection{Galactic white-dwarf binaries} \label{sec:GB}

For space-based detectors, these are the most numerous GW sources; they are also the only guaranteed source since several detectable systems (known as verification binaries) have already been identified by electromagnetic observations \citep{2006CQGra..23S.809S}.

Galactic binaries divide into two classes: the unresolvable and the resolvable galactic binaries. The unresolvable binaries overlap to form a stochastic background as discussed in section \ref{sec:stoch}. The distinction between resolvable and unresolvable is detector specific; here we choose LISA. This boundary shall not be too different for eLISA, but would move substantially for decihertz detectors. Plotted in \ref{app:a} with the label ``\emph{unresolvable galactic binaries}'' is the estimate of this background due to \citet{Nelemans} where an observation time of one year has been assumed,
\begin{equation}
h_\mathrm{c}(f)= 5\times 10^{-21} \left(\frac{f}{10^{-3}~\mathrm{Hz}}\right)^{-2/3} \; .
 \end{equation}
Estimates for the event rate of resolvable binaries centre around ${\mathcal{O}}(10^{3})$ events for eLISA. The boxes plotted in \ref{app:a} with the label ``\emph{resolvable galactic binaries}'' have a ratio $h_\mathrm{c}/h_{n}=50$ for eLISA at its peak sensitivity. The frequency range of the box is $\left(3\times10^{-4}\right.$--$\left.10^{-2}\right)~\mathrm{Hz}$, estimated from Monte Carlo population simulation results presented in \citet{Amaro-Seoane-et-al}.

\subsubsection{Extreme mass-ratio inspirals}

EMRIs occur when a compact stellar-mass object inspirals into a supermassive black hole. There is extreme uncertainty in the event rate for EMRIs due to the poorly constrained astrophysics in galactic centres \citep[e.g.,][]{Merritt2011}; the best guess estimate is around $25$ events per year with eLISA with SNR $\ge 20$ \citep{TheGravitationalUniverse}. The boxes labelled ``\emph{extreme mass ratio inspirals}'' plotted in \ref{app:a} have a characteristic strain of $h_\mathrm{c}=3\times 10^{-20}$ at $10^{-2}~\mathrm{Hz}$, which corresponds to a $10\Msun$ black hole inspiralling into a $10^{6}\Msun$ black hole at a luminosity distance of $1~\mathrm{Gpc}$. The frequency width of the box is somewhat uncertain; EMRI events can occur around a black hole of any mass, and hence EMRIs can, in principal, occur at any frequency. The boxes in \ref{app:a} are drawn with a width comparable to that of the LISA sensitivity curve.

\subsection{Sources for PTAs}

\subsubsection{Supermassive black hole binaries}

The main target for PTAs is a stochastic background of GWs produced by a population of supermassive black-hole binaries at cosmological distances \citep{SesanaVecchioColancino}. Supermassive black holes are known to lie at the centres of most galaxies and black-hole mergers are associated with the mergers of their host galaxies \citep{Volonteri2003,Ferrarese2005}. The current best 95\% confidence limit for the amplitude of the stochastic background is $h_\mathrm{c} = 2.7\times 10^{-15}$ at a frequency of $f_{0}=1~\mathrm{yr}^{-1}$ \citep{Shannon2013}. There is strong theoretical evidence that the actual background lies close to the current limit \citep{Sesana-2012}. The distribution of source amplitudes with frequency is dependent on the chirp mass distribution of binaries, which is astrophysically uncertain. Reflecting our lack of information about the true shape of this box, we draw it with a flat top in characteristic strain.

\subsubsection {Stochastic background of supermassive binaries}
Supermassive black-hole binaries at higher frequencies are inspiralling faster and, hence, there are fewer of them per frequency bin. At a certain frequency, these sources cease to be a background and become individually resolvable \citep{SesanaVecchioColancino,Sesana2009}. It is currently unclear whether PTAs shall detect an individual binary or a stochastic background first. Plotted in \ref{app:a} with the label ``\emph{stochastic background}'' is a third of the current limit with a cut off frequency of $f=1~\mathrm{yr}^{-1}$. This is suggested by Monte Carlo population studies \citep{SesanaVecchioColancino}, which give a range of plausible amplitudes the mean of which is plotted here. For the resolvable sources, labelled ``\emph{$\mathit{\approx 10^{9}}$ solar mass binaries}'', the amplitude of the current limit is plotted between $\left(3\times 10^{-9}\right.$--$\left.3\times 10^{-7}\right)~\mathrm{Hz}$.

\subsection{Cosmological sources}

In addition to the sources above, early Universe processes, such as inflation \citep{Grishchuk2005} or a first-order phase transition \citep{Binetruy2012}, could have created GWs. More speculatively, it has been hypothesised that cosmic strings could also be a potential source \citep{Damour2005,Binetruy2012,Aasi2014}. These relic GWs allow us to explore energy scales far beyond those accessible by other means, providing insight into new and exotic physics. The excitement surrounding the tentative discovery by BICEP2 of the imprint of primordial GWs (generated during inflation) in the cosmic microwave background \citep{Ade2014}, and the subsequent flurry of activity, has shown the scientific potential of such cosmological GWs. These GW signals are so alluring because they probe unknown physics; this also makes them difficult to predict. Cosmological stochastic backgrounds have been predicted across a range of frequencies with considerable variation in amplitude. As a consequence of this uncertainty, although we could learn much from measuring these signals, we have not included them amongst the sources in \ref{app:a}.

The sensitivity curves plotted in \ref{app:a} would change substantially for stochastic backgrounds. This is because searches for stochastic backgrounds utilise the cross correlation of the outputs of multiple detectors for detection \citep{1999PhRvD..59j2001A,2000PhR...331..283M}, instead of the individual outputs. Therefore, the sensitivity of a network of ground based detectors is much greater than that of any individual detector. The shape of the sensitivity curve for a stochastic background would also differ from the curves plotted here, for a discussion of this point see \cite{2013PhRvD..88l4032T}.

\section{Concluding remarks}\label{sec:discussion}

When quantifying the sensitivity of a GW detector and the loudness of a GW source, there are three commonly used quantities: the characteristic strain, the power spectral density, and the spectral energy density. We have carefully defined these quantities and derived the relationships between them. The characteristic strain (section \ref{sec:character-strain}) is most directly related to the SNR, the PSD is mostly closely related to the mean square amplitude in the detector, and the energy density has a clear physical interpretation. We have produced example plots using each of these quantities for a wide range of detectors and sources. The predicted source amplitudes are based on astrophysical estimates of the event rates and are subject to varying degrees of uncertainty.\footnote{There is also the exciting possibility of sources yet to be considered.} Interactive versions of these plots, with user-specified detectors and sources, are available on-line at \url{http://rhcole.com/apps/GWplotter}. Trying to summarise an entire field of astronomy on one plot is an impossible task; however, we hope that the figures and analysis presented here provide useful insight.

\ack{CJM, RHC and CPLB are supported by the Science \& Technology Facilities Council. We are indebted to Jonathan Gair, for useful advice and proof-reading the manuscript. We also thank Stephen Taylor and Justin Ellis for useful conversations about pulsar timing. The authors thanks a number of members of the LIGO--Virgo Scientific Collaboration, especially Chris Messenger, for useful comments and suggestions. This paper has been assigned LIGO document LIGO-P1400129.}

\appendix
\section{Sensitivity curves}\label{app:a}
The plots in this section show all of the detectors and sources described in the main text. Clearer, interactive versions of these plots, allowing for removal of any of the curves, may be created and downloaded on-line, \url{http://rhcole.com/apps/GWplotter}. The detector noise curves all have their resonance spikes removed for clarity. 

\begin{figure}[h!]
 \centering
 \includegraphics[trim=0cm 0cm 0cm 0cm, width=0.8\textwidth]{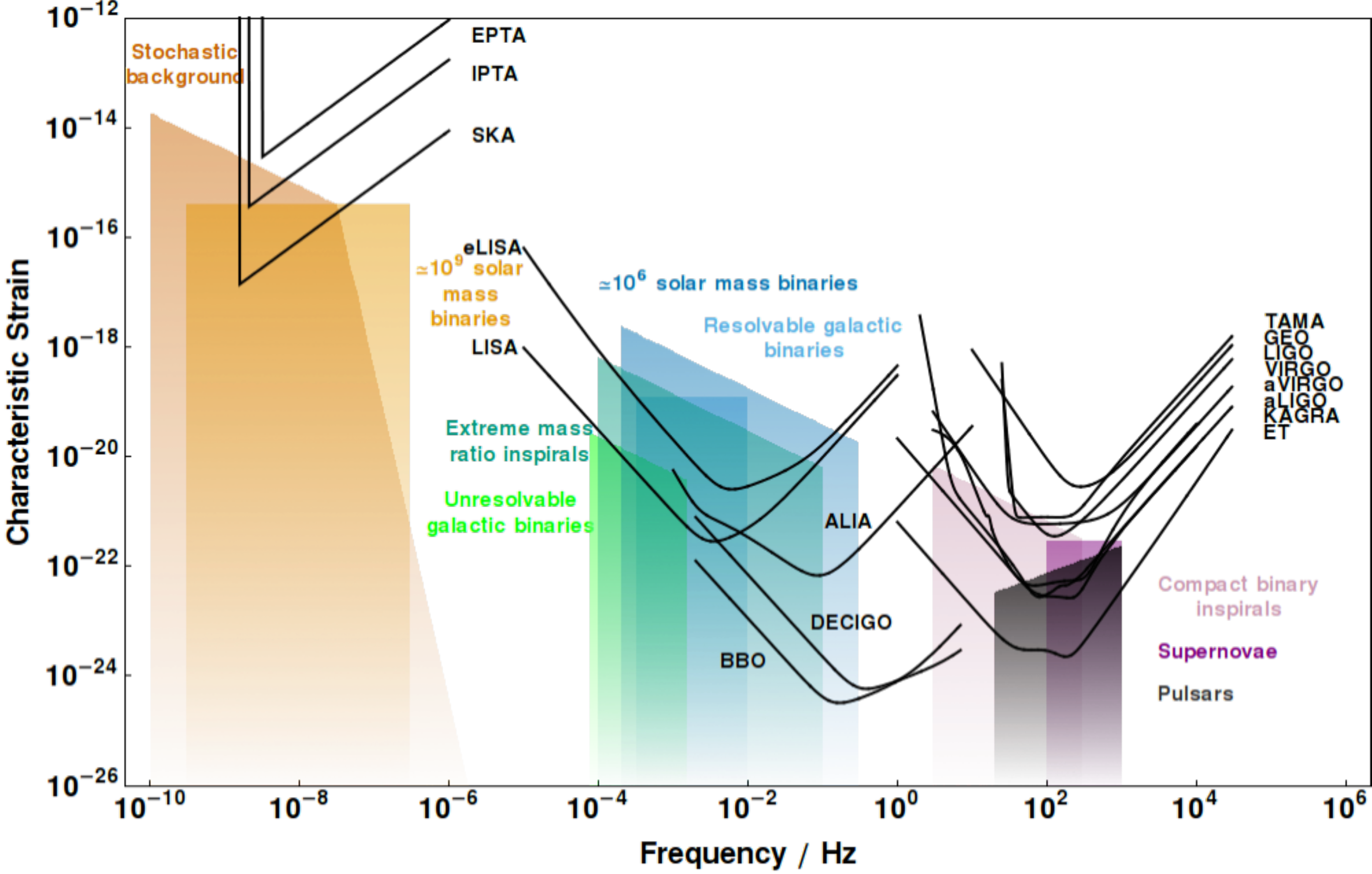}
 \caption{A plot of characteristic strain against frequency for a variety of detectors and sources.}
 \label{fig:hc}
\end{figure}

\begin{figure}
 \centering
 \includegraphics[trim=0cm 0cm 0cm 0cm, width=0.8\textwidth]{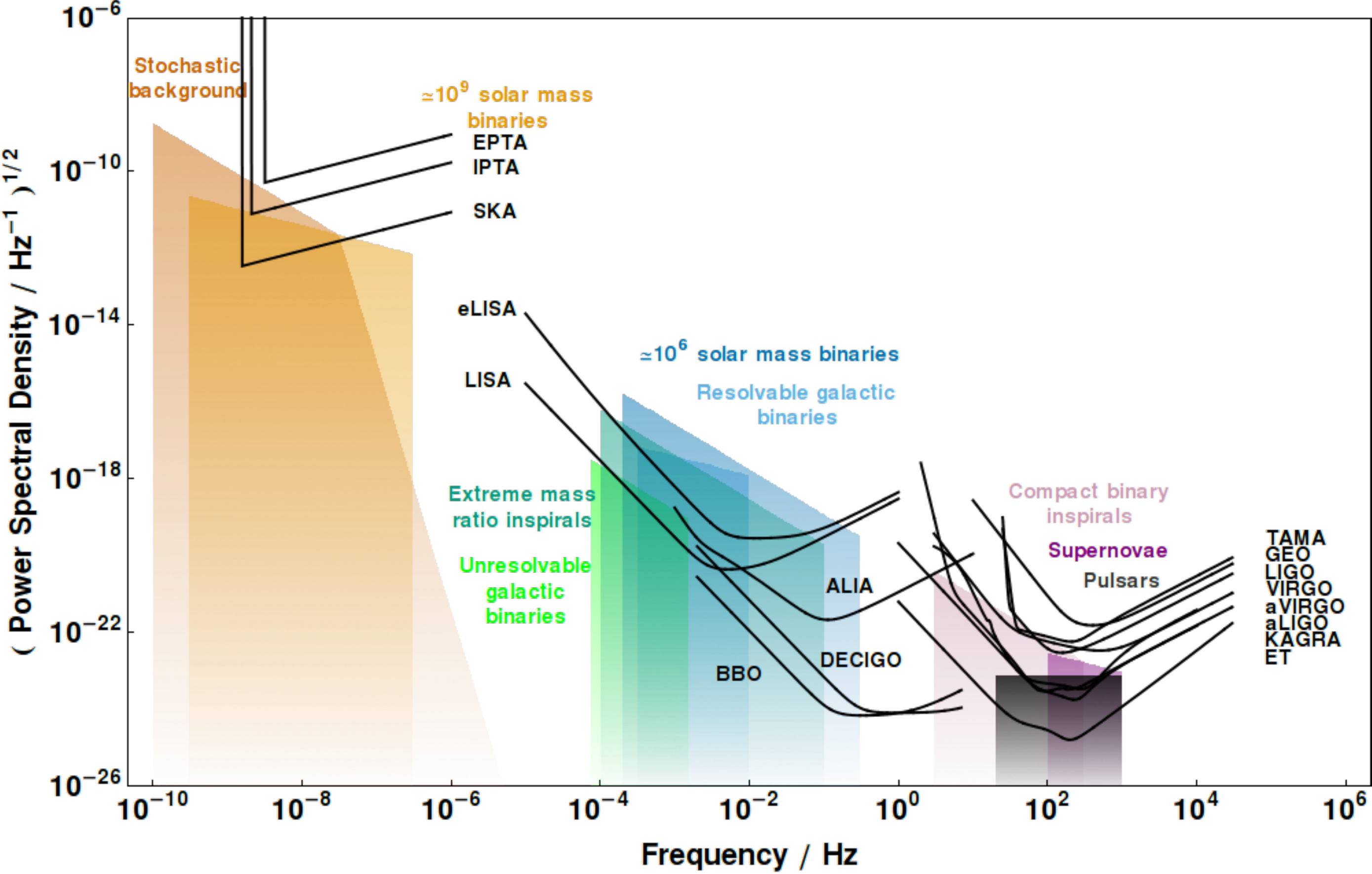}
 \caption{A plot of the square root of power spectral density against frequency for a variety of detectors and sources.}
 \label{fig:S}
\end{figure}

\begin{figure}
 \centering
 \includegraphics[trim=0cm 0cm 0cm 0cm, width=0.8\textwidth]{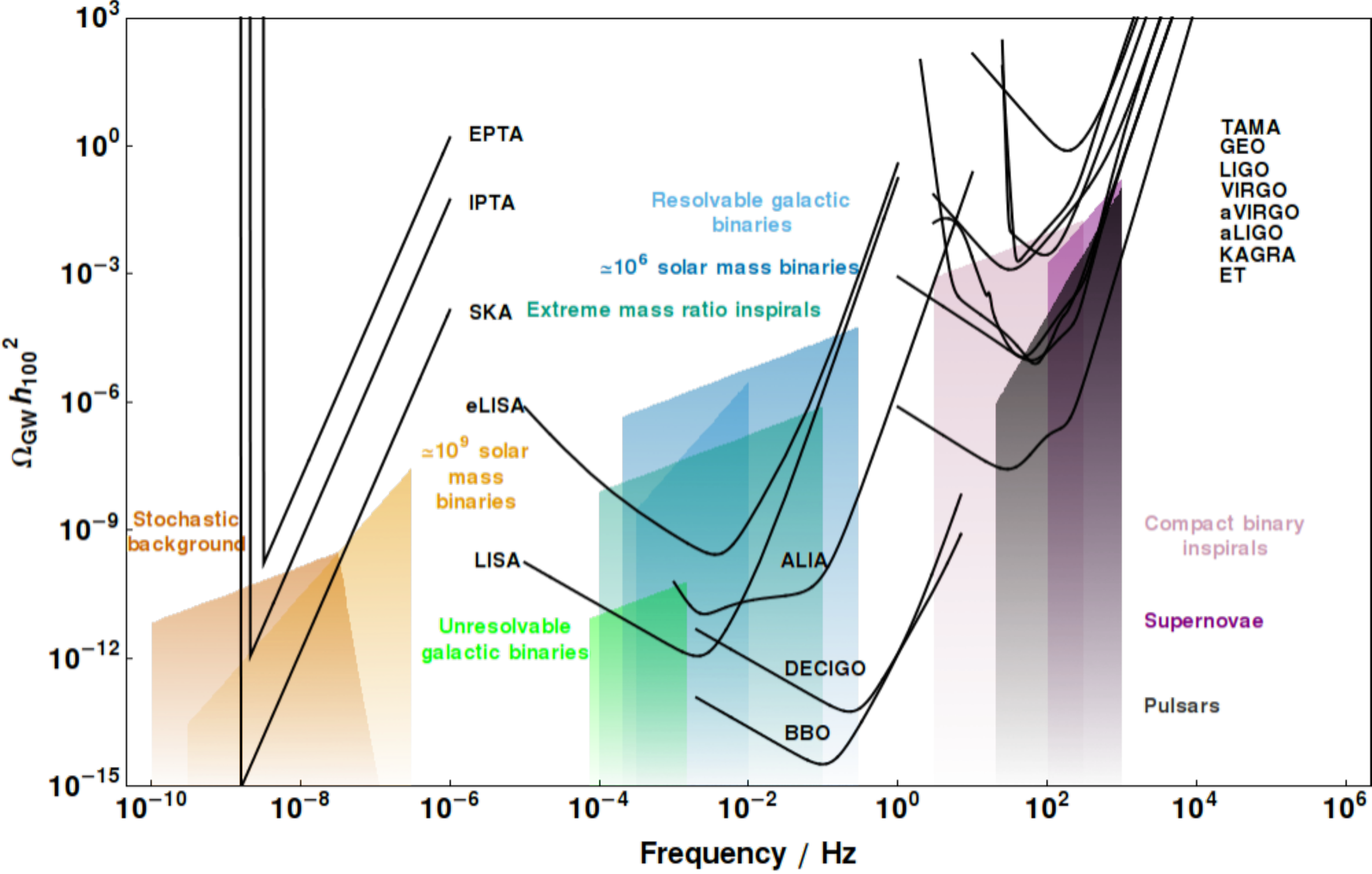}
 \caption{A plot of the dimensionless energy density in GWs against frequency for a variety of detectors and sources.}
 \label{fig:omega}
\end{figure}

\clearpage

\bibliographystyle{jphysicsCB}
\bibliography{bibliography}

\end{document}